\documentclass[journal=jacsat,manuscript=article]{achemso}

\usepackage[version=3]{mhchem} 
\usepackage{hyperref}
\usepackage{amsmath}
\usepackage{amssymb}

\author{Eline K. Kempkes}
\author{Alberto~Pérez de Alba Ortíz}
\email{a.perezdealbaortiz@uva.nl}
\affiliation[UvA]
{Computational Soft Matter Lab, Computational Chemistry Group and Computational Science Lab, Van ’t Hoff Institute for Molecular Sciences and Informatics Institute, University of Amsterdam, Science Park 904, 1098 XH Amsterdam, The Netherlands.}

\title
  {Bayesian umbrella quadrature accelerates free-energy calculations across diverse molecular systems and processes}

\abbreviations{IR,NMR,UV}

\begin{document}







\begin{abstract}
Biased sampling in molecular dynamics simulations overcomes timescale limitations and delivers free-energy landscapes, essential to understand complex atomistic phenomena. However, when applied across diverse systems and processes, biasing protocols often require time- and resource-consuming fine-tuning. In search for robustness, we boost a prominent biasing method—Umbrella Sampling. To estimate the value of an integral, i.e., the free energy, our Bayesian Umbrella Quadrature (BUQ) method iteratively selects gradient samples, i.e., bias locations, that most reduce the posterior integral variance based on a noise-tolerant Gaussian process model, which also effectively interpolates between samples. We validate the method for a conformational change in a small peptide, a water-to-ice phase transition, and a substitution chemical reaction; obtaining excellent accuracies and speedups. To ease adoption of this more automated and universal free-energy method, we interface BUQ with wide-spread simulation packages and share hyperparametrization guidelines. 
\end{abstract}

\section{Introduction}

Accurate calculation of free-energy landscapes is central to understanding and predicting chemical reactions and conformational changes in molecules as well as phase transitions in materials. These landscapes are projected onto key degrees of freedom for the system and process at hand---i.e., reaction coordinates, collective variables (CVs), or order parameters, which we will refer to as CVs without loss of generality. Free energies quantify the occurrences, i.e., probabilities, of metastable states and the transition rates and mechanisms between them. Since transitions of interest often occur on timescales inaccessible to standard molecular dynamics (MD) simulations~\cite{Alder1959}, enhanced sampling methods have been introduced.~\cite{Henin2022,Mohr2024}. Among these, biasing approaches such as steered molecular dynamics~\cite{Jarzynski1997,Grubmller1996} or metadynamics~\cite{laio2002,Bussi2020} accelerate sampling by applying external potentials that guide the system toward infrequently sampled regions. One of the most well-established biasing methods is Umbrella Sampling (US),\cite{Torrie1977} with over 400 citations per year since 2019 and growing, according to Google Scholar. The main idea of US is to use a series of biased simulations, or windows, each restraining the system around a specific region of the $d$-dimensional CV-space~\cite{kastner2011}. In standard US, a typical biasing potential for the $i$-th window, acting on the $j$-th CV, is defined as:
\begin{equation}
V_{\text{bias}}^{(i)}(s_j) = \frac{1}{2} \kappa_j^{(i)} \left(s_j -\hat{s}^{(i)}_j\right)^2 . 
\end{equation}

\noindent For simplicity, it is assumed that the CVs, $\mathbf{s}=(s_1,\dots,s_d)$, are independent, however, this can be extended to dependent CVs~\cite{kastner2011, kastner2009} (see Section \hyperref[sec:methods]{Methods}). $\hat{s}^{(i)}_j$ denotes the bias center for window $i$, for CV $s_j$, with a harmonic force constant $\kappa_j^{(i)}$.  Using multiple biased simulations at different values of $\mathbf{\hat{s}}^{(i)}$, relevant regions of CV-space can be explored.
The free-energy landscape can be reconstructed using the Weighted Histogram Analysis Method (WHAM), which works by combining bias-corrected histograms from multiple windows~\cite{Kumar1992}. US data can also be processed by the Multistate Bennett Acceptance Ratio (MBAR) method, which works by statistically reweighting samples from multiple thermodynamic states to estimate free energies~\cite{Shirts2008}. A limitation of both approaches is that the sampled distributions from adjacent umbrellas must overlap sufficiently, and in WHAM, the choice of bin size directly influences the estimated distributions. An alternative is Umbrella Integration (UI)~\cite{kastner2005}, where we use the negative of the ensemble-averaged biasing force to approximate the free-energy gradient for window $i$:
\begin{equation}
    \frac{\partial A(s_j)^{(i)}}{\partial s_j}\approx 
     - \frac{\partial V_\text{bias}(s)^{(i)}}{\partial s_j}
    = -\kappa_j^{(i)}\big(\overline{s}_j^{(i)}-\hat{s}^{(i)}_j\big). 
\end{equation}
Afterwards, numerical integration can be used to recover the full free-energy surface. Note that this approximation requires a unimodal distribution of the ensemble-averaged biasing force at each window.
Here, as in done frequently in literature, we refer to the resulting potential of mean force (PMF) as a free energy, but we note that the exact relation between the two depends on the selection of CVs and contribution of the Jacobian.\cite{wong2012exact} Compared to WHAM and MBAR, UI substantially reduces errors associated with histogram discretization and does not require window overlap\cite{kastner2005,kastner2006}. Moreover, UI is easy to automate, requiring only averaging the biasing force and verifying its unimodality. A disadvantage is that the gradient of the free energy might change at an unsampled region, resulting in an inaccurate free-energy calculation.

The challenge in UI, US and restrained MD methods in general lies in the careful tuning of the parameters, e.g., the number, strength (i.e., $\kappa_j^{(i)}$), and positions ($\mathbf{\hat{s}}^{(i)}$) of the biasing potentials. This can be challenging for complex systems, especially those involving more than one CV. Moreover, for different processes---e.g., protonation reactions, ligand dissociation, homogeneous nucleation---these settings can be extremely different, since they correspond to fundamentally different phenomena and interactions---e.g., covalent bonds, hydrogen bonds, van der Waals forces, electrostatic repulsion or attraction, etc. This makes the tuning process typically system-specific and highly non-trivial, since unknown barrier heights, metastable valleys and other landscape features must be resolved~\cite{kastner2006,pietrucci2017}. This prevents generalization and performance optimization across diverse systems, since poorly chosen biasing parameters can lead to undersampled regions and to wasted computational efforts. Even when dealing with similar systems, small modifications---e.g., changes in pH, the presence of an ion, or a change in temperature---can reshape the free-energy landscape in such way that a former biasing protocol becomes ineffective.

In other words, UI and US protocols can become inefficient and inaccurate when applied across diverse systems. To overcome these limitations, several adaptive schemes have been proposed. Early approaches introduced feedback-based optimization of biasing potentials and window placement to enhance sampling efficiency~\cite{Wojtas-Niziurski2013,Shirts2020,mezei1987,bartels1997,bartels1998}. More recent developments have incorporated self-learning frameworks that adjust these parameters on the fly ~\cite{Novack2025,Das2025}, and used machine-learning-based approaches to further generalize this idea~\cite{Mitsuta2024,Mitsuta2025}.
Recently, Gaussian process regression has also been successfully applied to reconstruct free-energy surfaces from US and other simulation data, enabling smooth free-energy surface estimation with quantified uncertainty \cite{Ladygin2021,Mones2016,Stecher2014}. Related to this, Bayesian inference and related probabilistic approaches have been successfully applied to free-energy estimation, structure refinement, and parameter optimization in molecular systems. Bayesian methods allow for principled treatment of uncertainties, heterogeneous states, and noisy or sparse data. Representative examples include  extracting free energies or transition rates from cryogenic electron microscopy data \cite{Giraldo-Barreto2021},  path sampling data~\cite{Terrier2015}, equilibrium and non-equilibrium data, \cite{Hummer2005, Habeck2012,Maragakis2008}, etc.

\begin{figure}[ht]
\centering
\includegraphics[width=\linewidth]{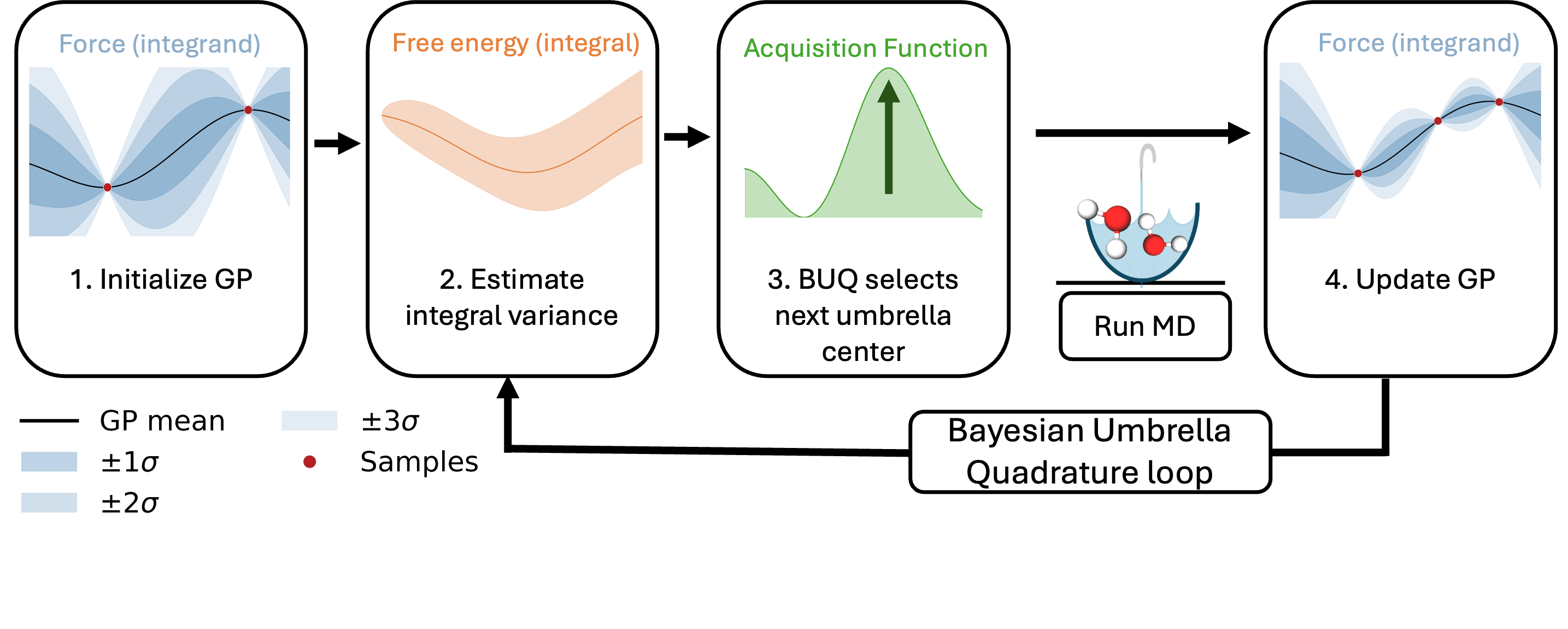}
\caption{Schematic representation of the Bayesian Umbrella Quadrature (BUQ) loop applied to free-energy calculations. (1) The Gaussian Process (GP) gradient model is initialized considering a variance $\sigma$ and a few initial gradient samples from umbrella-biased molecular dynamics (MD) simulations. (2) Based on the GP gradient model, the variance of the integral, i.e., the free energy, is estimated. (3) Using an acquisition function, BUQ selects the next umbrella center at which sampling the gradient will maximally reduce the variance of the free-energy estimate. (4) A new biased MD simulation is run at the proposed umbrella center, generating a new gradient sample. This new observation is incorporated into the GP model, and the loop is repeated until convergence. This adaptive strategy allows BUQ to prioritize the most informative sampling regions, reducing the computational cost compared to standard grid-based umbrella integration.}
\label{fig:bqmethod}
\end{figure}

Inspired by the rise of successful Bayesian approaches, here we propose a Bayesian Umbrella Quadrature (BUQ) framework as a general and automated solution for UI-based free-energy calculations. Bayesian Quadrature (BQ) is an established method for estimating the value of integrals based on noisy samples of the gradient, leveraging Bayesian inference~\cite{OHagan1991,gunter2014}. To our knowledge, this is the first time BQ is integrated into an enhanced sampling MD pipeline. BUQ integrates BQ in a UI framework, where we estimate the value of an integral, i.e., the free-energy landscape, based on iteratively selecting samples of its gradient, i.e., the negative of the force exerted by the umbrellas. BQ uses a Gaussian Process (GP), i.e., a noise-tolerant probabilistic model, to estimate the value of the gradient, which can be numerically integrated afterwards. Figure~\ref{fig:bqmethod} illustrates the iterative process. Starting from either prior information about the free-energy gradient or from a few initial gradient samples, BUQ determines the next \textit{most informative} point to sample the gradient, i.e., the location of the next bias potential, $\mathbf{\hat{s}}^{(i)}$. Here, the definition of "most informative" is given by an acquisition function, which can be focused on reducing the overall uncertainty of the free energy, i.e., Integral Variance Reduction (IVR), or steered to search for free-energy minima or maxima. This iterative process is repeated such that the integral of the GP gradient surrogate converges to a stable estimate of the free-energy surface (see Section \hyperref[sec:methods]{Methods}). 

The two main strengths of BUQ are: (1) the GP interpolates the gradient, providing both an accurate estimate and an associated uncertainty, and (2) the acquisition function guides sampling toward the most informative regions, thereby reducing overall computational effort. This directly addresses the limitations of UI,  reducing the risk of undersampling key regions and improving the interpolation of the gradient between samples. An additional strength of BUQ is its ability to integrate existent knowledge, e.g., by integrating data from previous US runs or an initial guess into the GP.

We demonstrate the generality and efficiency of our BUQ method on three benchmark problems of varying complexity. First, we benchmark the performance of BUQ against standard UI at calculating the conformational free energy of alanine dipeptide in vacuo projected onto the $\phi$ and $\psi$ dihedral angles, showing BUQ converges 1.6 times as fast. Next, to show the robustness of BUQ when handling a collective process, we calculate the nucleation free-energy profile of the phase transition from water to ice Ih, observing superior BUQ efficiency with respect to modern enhanced sampling methods. Finally, we tackle a different challenge: predicting the barrier of an S$_\text{N}2$ chemical reaction, where bond breaking introduces fundamentally different interactions, and thus a different free-energy landscape. With these three examples, we show that one of the key strengths of BUQ lies in its generalizable performance; BUQ can be applied across a wide range of systems and processes, under diverse conditions. We also show that BUQ hyperparametrization is of comparable difficulty to existent methods, and provide pointers to facilitate it. To ease the adoption of BUQ, we provide an open-source implementation interfaced with well-established packages for enhanced sampling---i.e., PLUMED~\cite{consortium2019,Tribello2014}—and for MD of different systems---GROMACS~\cite{Berendsen1995}, LAMMPS~\cite{Thompson2022} and ASE~\cite{HjorthLarsen2017}.  All of this considered, BUQ promises a more universal and automated approach to evermore relevant free-energy calculations of molecules and materials.

\section{Methods}

\label{sec:methods}
We begin this section by outlining the standard umbrella integration (UI) approach. Next, we discuss the theory of Bayesian Quadrature (BQ) in the context of free-energy estimation, focusing on its two core components: Gaussian Processes (GP) and acquisition functions. Finally, we describe the integration of UI in BQ, i.e. Bayesian Umbrella Quadrature (BUQ). 

\subsection*{Umbrella Integration}

Umbrella sampling (US) probes free-energy landscapes projected onto a set of key descriptive degrees of freedom---i.e., collective variables (CVs),  $\mathbf{s}=(s_1,\dots,s_j)$---using harmonic bias potentials. A typical bias at window $i$ for multiple CVs takes the form~\cite{kastner2011, kastner2009}:
\begin{equation}
V_{\text{bias}}^{(i)}(\mathbf{s}) = \frac{1}{2}  (\mathbf{s} -\hat{\mathbf{s}}^{(i)})^\dagger \mathbf{U} (\mathbf{s} -\hat{\mathbf{s}}^{(i)}), 
\end{equation}
\noindent where $\hat{\mathbf{s}}^{(i)}$ denotes the bias center for window $i$, and $\mathbf{U}$ denotes the matrix of umbrella force constants.  Using multiple biased simulations at different values of $\mathbf{\hat{s}}^{(i)}$, relevant regions of CV-space can be explored. Multiple biased simulations are carried out in a number of windows. In each biased simulation, an averaged biasing force can be calculated based on the average CV vector in window $i$. Given large enough force constants and unimodal force distributions, each of these average biasing forces can be used to approximate the negative gradient of the underlying free energy at the bias center. The use of the covariance can be avoided by truncation of the Jacobian terms \cite{kastner2009}. This yields the approximation:  
\begin{equation}
\nabla A(\mathbf{s})^{(i)} |_{{\mathbf{s}=\hat{\mathbf{s}}^{(i)}}} \approx - \mathbf{U} \dot (\langle \mathbf{s}^{(i)}\rangle - \hat{\mathbf{s}}^{(i)} ).
\end{equation}
In practice, one collects these gradient estimates from all windows and numerically integrates them to recover the free-energy surface $A(\mathbf{s})$.  Thus, UI provides an effective way to compute $\nabla A$ from biased simulations, after which numerical integration yields $A(\mathbf{s})$.

\subsection*{Bayesian Quadrature}
Bayesian quadrature treats the integral of an expensive function as a Bayesian inference problem. Specifically, it does this by modeling the integrand with a Gaussian Process (GP), and refining this surrogate through sequentially selected evaluations of the gradient.~\cite{OHagan1991}
In our application, we aim to compute the free energy with respect to a reference, $ A(\mathbf{s}_0)$, which can be written formally as a path integral of the mean biasing force:
$$A(\mathbf{s}) = A(\mathbf{s}_0) + \int_{\mathbf{s}_0}^{\mathbf{s}} -\mathbf{F}(\mathbf{s'})\cdot d\mathbf{s'},$$
\noindent where $-\mathbf{F}(\mathbf{s})$ is the negative mean force, i.e.,the integrand. For simplicity, we set $ A(\mathbf{s}_0) = 0$. We model the unknown integrand $-\mathbf{F}(\mathbf{s})$ with a Gaussian process (GP) prior:
\begin{equation}
    -\mathbf{F}(\mathbf{s}) \sim \mathcal{GP}\bigl(\mathbf{m}(\mathbf{s}),\mathbf{K}(\mathbf{s},\mathbf{s}')\bigr),
\end{equation}
\noindent where $\mathbf{m}$ is the prior mean, often taken equal to zero, and $\mathbf{K}$ is a matrix-valued covariance kernel. In the independent-CV case, one can use independent GPs for each force component. We use standard stationary kernels---e.g. Radial Basis Function (RBF) or Matérn-$\nu$ kernels ~\cite{Rasmussen2006}---to encode smoothness: the RBF kernel implies an infinitely differentiable integrand, whereas Matérn kernels allow for tunable roughness---e.g.\ Matérn-1/2 permits non-differentiable, rough functions, while Matérn-5/2 yields a smoother function. For a more detailed discussion, see the Supporting Information.  
This surrogate GP is then updated with $n$ initial samples $D={(\mathbf{s}_i,-\mathbf{F}(\mathbf{s}_i))}_{i=1}^n$ from biased MD runs. Each output dimension, i.e, CV, $j$ has its own mean $m_{D}(s_j)$ and covariance $k_{D}(s_j,s_j')$, given by:
\begin{align}
    m_{D}(s_j) = m(s_j) + k(s_j,S)[K(S,S)]^{-1}\bigl(y-m(S)\bigr), \\
    k_{D}(s_j,s_j') = k(s_j,s_j') - k(s_j,S)[K(S,S)]^{-1}k(S,s_j'),
\end{align}
where $S=[s_{j_1},\ldots,s_{j_n}]$, i.e., samples of $s_j$, $y=[-F(s_{j_1}),\dots,-F(s_{j_n})]^\top$, and $k$ is the scalar kernel for each component. 

Once the posterior is obtained, we can integrate the surrogate analytically to estimate the free energy.  In particular, the expected value of the integral is obtained by integrating the GP posterior mean:
 \begin{equation}
      \mathbb{E}[A]=\int_{l_1}^{u_1}\cdots\int_{l_d}^{u_d} m_D(s_1,\dots,s_d)\,ds_1\cdots ds_d.
 \end{equation}
We integrate over finite ranges of the CVs (with lower bounds $l_1,...,l_d$, and upper bounds $u_1,...,u_d$).
The posterior integral variance of CV $s_j$ can be obtained by integrating the covariance:
\begin{equation}
    \mathbb{V}[A] = \iint k_D(s_j,s'_j)ds_jds_{j}'.
\end{equation}
Thus, we model the force function by a GP, giving both a point estimate and uncertainty for $A$.
We improve efficiency by actively selecting new sampling points via an acquisition function. A common choice in BQ is the integral-variance-reduction (IVR) criterion, which quantifies how much the posterior variance of the integral would drop by sampling at a candidate point~\cite{gessner2020}. Formally, if $\mathbb{V}n$ is the current variance of $A$, then the IVR score for a new point $\mathbf{s}$ is:
\begin{equation}
    a_{\rm IVR}(\mathbf{s}) = \mathbb{V}n -\mathbb{V}\bigl[A\mid D\cup{\mathbf{s}}\bigr],
\end{equation}
i.e. the reduction in uncertainty by sampling this point. One can also express this as a normalized squared-correlation $\rho^2(\mathbf{s})\in[0,1]$ between the integrand at $\mathbf{s}$ and the integral. To bias sampling toward low-free-energy regions, we combine IVR with an exploitation term based on the GP predictive mean. Let $a_{\rm FES}(\mathbf{s})$ be the (normalized) GP estimate of $A(\mathbf{s})$ based on $D$, then we define:
\begin{equation}
    a_{\rm total}(\mathbf{s}) = -\lambda a_{\rm FES}(\mathbf{s}) + (1-\lambda)a_{\rm IVR}(\mathbf{s}),
\end{equation}
where $\lambda\in[0,1]$ balances exploration (variance reduction) against exploitation (favoring lower predicted $A$). This hybrid strategy is analogous to combined exploration-exploitation in Bayesian optimization. For a more detailed discussion, we refer to the Supporting Information.

At each iteration, we maximize the chosen acquisition $a_{\rm total}(\mathbf{s})$ over $\mathbf{s}$ to select the next sample point $\mathbf{s}_{\rm new}$. We then perform a restrained MD (US) simulation at $\mathbf{s}_{\rm new}$ to measure the new mean force $-\mathbf{F}(\mathbf{s}_{\rm new})$, add this to the dataset $D$, and update the GP to obtain a new posterior mean, i.e. free-energy gradient  prediction. In summary, the iterative active BUQ procedure is:
\begin{enumerate}
    \item Initialization: Choose an initial set of $\tilde{n}$ samples in CV-space and run biased MD at each of them to obtain forces,    $D_{\text{initial}}= {(\mathbf{s}_i,-\mathbf{F}(\mathbf{s}_i))}_{i=1}^{\tilde{n}}$. Fit a GP (kernel and hyperparameters) to this data.
    \item Inference: Compute the GP posterior mean $\mathbf{m}_D(\mathbf{s})$ and covariance $\mathbf{K}_D(\mathbf{s},\mathbf{s}')$, then integrate to get $\mathbb{E}[A]$ and $\mathbb{V}[A]$. Evaluate the acquisition $a(\mathbf{s})$ (e.g., IVR or combined).
    \item Acquisition: Select the next point $\mathbf{s}_{\rm new} = \arg\max{\mathbf{s}}_{ a(\mathbf{s})}$. Run a new restrained MD at $\mathbf{s}_{\rm new}$ to obtain $-\mathbf{F}(\mathbf{s}_{\rm new})$.
    \item Update: Augment the dataset with $(\mathbf{s}_{\rm new},-\mathbf{F}(\mathbf{s}_{\rm new}))$, refit the GP, and repeat steps 2–3 until convergence (e.g., variance is small or landscape does not change significantly).
\end{enumerate}
This loop systematically leverages both the physical insight of UI (averaged biasing forces as approximations to free-energy gradients) and the sample-efficiency of Bayesian quadrature to accelerate free-energy estimation.

\section{Results}

\subsection*{Conformational change in alanine dipeptide}
\begin{figure}[ht]
\centering
\includegraphics[width=\linewidth]{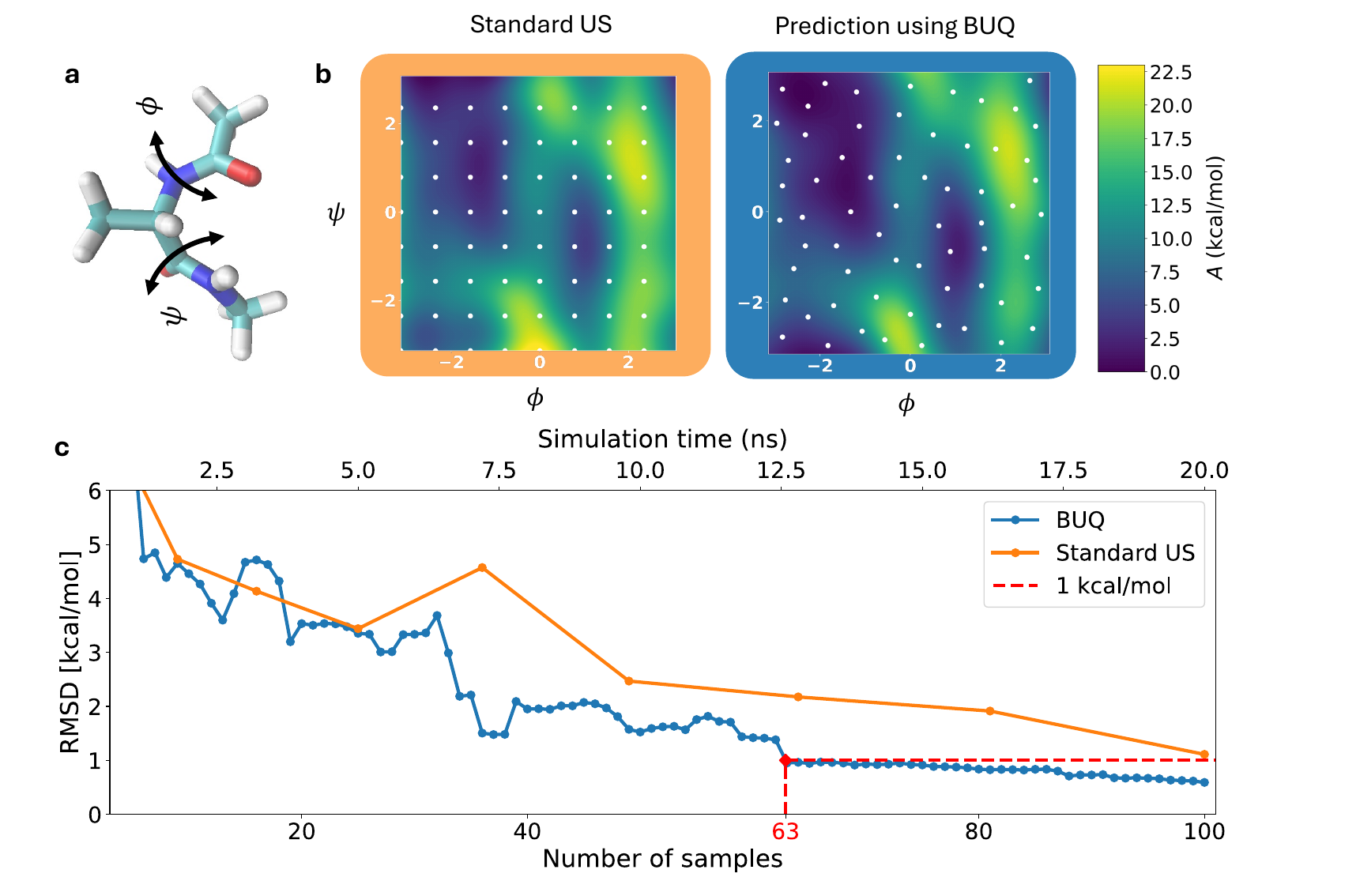}
\caption{Bayesian Umbrella Quadrature (BUQ) sampling of the conformational free-energy landscape of alanine dipeptide. \textbf{a}, Representation of the alanine dipeptide molecule with relevant collective variables (CVs), the dihedral angles  $\phi$ and $\psi$. \textbf{b}, Conformational free-energy landscape of alanine dipeptide in vacuo obtained by BUQ (blue) compared with a standard umbrella integration (UI) (orange) benchmark.  The distribution of umbrella centers shows how BUQ adaptively places samples along the contours of valleys and barriers on the free-energy landscape. \textbf{c}, Root-mean-squared deviation (RMSD) of BUQ and UI with respect to the ground truth as a function of the number of samples, or total simulation time.}
\label{fig:adipep}
\end{figure}

To demonstrate the performance of our BUQ framework, we start with a conformational transition in a biomolecular system traditionally used to illustrate rare-event sampling, namely, an alanine dipeptide molecule~\cite{laio2002,bolhuis2000}. We focus on the free-energy landscape defined by alanine dipeptide's $\phi$ and $\psi$ dihedral angles (see Fig.~\ref{fig:adipep}\textbf{a}). To measure the performance of BUQ, we consider a two-folded metric. First, we compare the free-energy surface calculated by our BUQ method after the $i$-th query, $A_{i,BUQ}$, to a thoroughly converged metadynamics reference, $\hat{A_i}$, that is considered the ground truth (see Supporting Information for details). We do this comparison by computing the grid-based Root Mean Squared Deviation (RMSD) of the BUQ free-energy surface, $A_{i,BUQ}$, with respect the metadynamics reference, $\hat{A_i}$ (see Supporting Information for details). A lower RMSD indicates a closer match to the benchmark. In general, a fast and consistent decrease of RMSD with the number of queries is desirable. \\Second, to compare the performance our BUQ method to standard UI, we perform UI on uniform grids of various resolutions, each with a different amount of $l = m \times m$ umbrellas, yielding a free-energy surface $A_{l,UI}$. This allows for a fair comparison of MD computational costs, as calculating a free-energy surface $A_{i,BUQ}$ requires the same MD steps as calculating a surface $A_{l,UI}$ for $l  = i$. Note that for both the BUQ and the standard UI free-energy calculation we use a GP to interpolate the free-energy gradient, so it is only the acquisition of points that is evaluated by this benchmark, see Fig.~\ref{fig:adipep}\textbf{b} and see Methods for more details.

In Fig.~\ref{fig:adipep}\textbf{c}, we present BUQ sampling results compared against the aforementioned benchmarks. BUQ was performed with a Matérn-$5/2$ kernel with a lengthscale of 0.75 rad and a weight for exploiting the free-energy minima $\lambda = 0.1$ (see Methods). First, we highlight that, considering the number of queries, BUQ sampling achieves a faster approximation of the ground truth free energy than standard UI. A full UI with a $10\times 10$ umbrella grid yields an RMSD of  $\sim$1.1 kcal/mol relative to the ground truth. Remarkably, BUQ surpasses this accuracy with only 63 samples, i.e. , 63\% of the MD cost, as shown in Fig~\ref{fig:adipep}. In fact, also after only 63 samples ($\approx$12.6 ns total sampling) BUQ achieves an RMSD <1 kcal/mol with respect to the ground truth, a level that is often considered chemical accuracy \cite{Pople1999}. This yields a 1.6 sample-efficiency speedup for BUQ. For most of the run and for all queries after query 25, BUQ’s RMSD is consistently lower than that of UI for the same number of samples. Thanks to the IVR acquisition function, the BUQ sample points tend to follow the contours of valleys and barriers in the free-energy landscape, focusing MD effort on the most informative regions. These results show that BUQ can accurately calculate the conformational free-energy landscape of a small biomolecule with faster convergence than standard UI. Similar, robust performance is also achieved with variations of the BUQ kernel, hyperparameters and acquisition function, as we show in the Supporting Information.

\subsection*{Phase transition from water to ice}
To demonstrate BUQ on a nucleation process, we study the free-energy profile for the phase change from liquid water to ice Ih. The freezing of water has been a popular, relevant and challenging case study for enhanced sampling methods \cite{Sun2024,Conde2017,GarcaFernndez2006}. Following Piaggi and Car \cite{Piaggi2020},  we use the TIP4P/Ice model and the environment similarity order parameter~\cite{piaggi2019} to distinguish between liquid- and ice-like water molecules. Put simply, the environment similarity is a measure of how closely the local neighbor arrangement of a specific atom matches a given template, e.g., that of ice Ih. In practice, we bias the number of water oxygen atoms with an environment similarity higher than a certain threshold with respect to ice Ih, which we label $N_{\text{ice}}$. For more details see the Supporting Information. 

\begin{figure}[ht]
\centering
\includegraphics[width=0.9\linewidth]{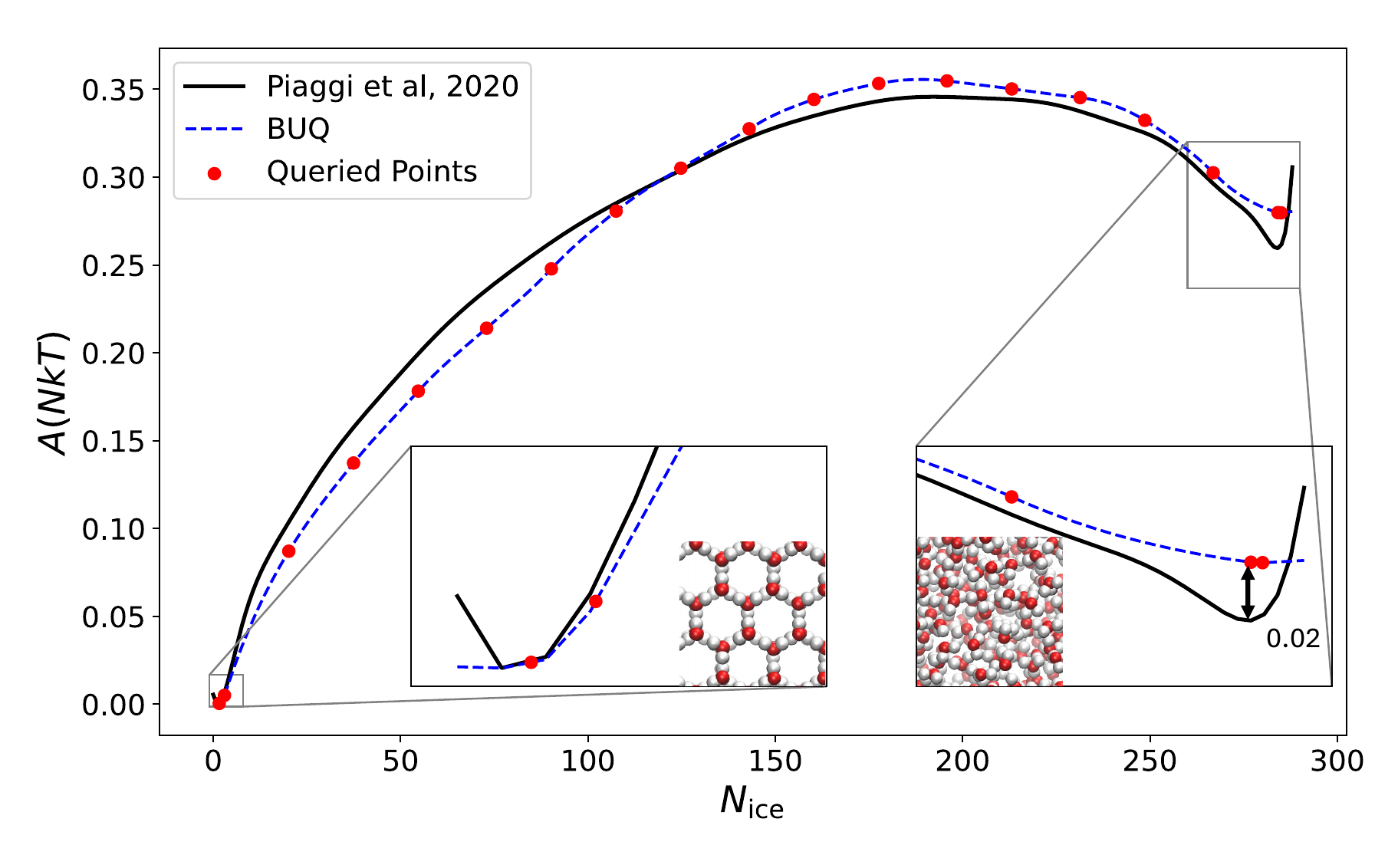}
\caption{Bayesian Umbrella Quadrature (BUQ) sampling of the nucleation free-energy profile of the phase transition from water to ice Ih. Free energy as a function of the number of ice-like water molecules, $N_\text{ice}$, obtained by BUQ (blue) compared against the reference profile from variational enhanced sampling (black) taken from Ref. \cite{Piaggi2020}. BUQ samples are indicated by red dots. }
\label{fig:phasetrans}
\end{figure}

In Fig. \ref{fig:phasetrans}, we show BUQ sampling results compared against variational enhanced sampling (VES) \cite{valsson2014}, as reported in Ref. \cite{Piaggi2020}. BUQ was performed with a Matérn-$1/2$ kernel with a lengthscale of 20 $N_\text{ice}$ and a weight for exploiting the free-energy minima $\lambda = 0.1$ (see Methods). As initialization, we run four samples near the stable states (see Supporting Information for details). After 15 production samples, we find that the predicted free-energy profile closely resembles the one obtained via VES, with a difference in barrier height of 0.01 $NkT$, with $N$ being the number of particles, $k$ the Boltzmann constant, and $T$ the temperature. The difference in the free-energy difference from the water to the ice state is 0.02 $NkT$. This small discrepancy can be explained by the dynamic nature of VES, where every walker is free to explore the whole CV range, compared to the static nature of BUQ sampling, where every sample is restrained. The locations of the two stable states in the $N_\text{ice}$ order parameter range also align well with previous work. In other words, BUQ reproduces both the barrier and the stable states correctly. This shows that BUQ can sample a phase transition involving anisotropic molecules. In total, we run 19 x 15 ns = 285 ns. In comparison, Ref.~\cite{Piaggi2020} used 4 walkers, where each walker ran for 200 ns, giving 800 ns in total. This gives BUQ a remarkable 2.8 speedup considering the number of MD timesteps. We note that the BUQ protocol required one rerun due to a failed sample (see Supporting Information), but this also shows how a sample can easily be corrected within the BUQ framework. 

\subsection*{S$_\text{N}2$ chemical reaction between ethyl chloride and a fluoride anion}
\begin{figure}[ht]
\centering
\includegraphics[width=\linewidth]{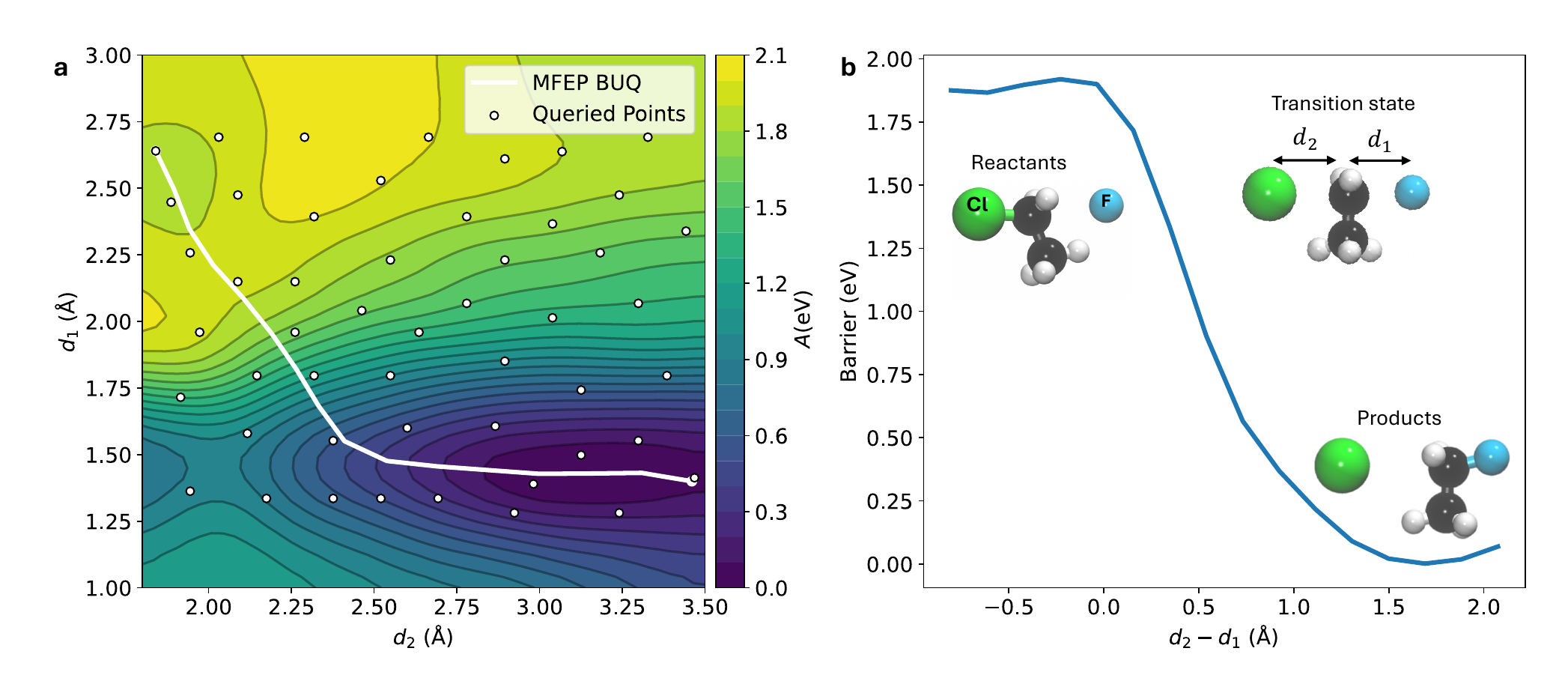}
\caption{Bayesian Umbrella Quadrature (BUQ) sampling of the S$_\text{N}2$ reaction free-energy landscape between ethyl chloride and a fluoride anion. \textbf{a}, The free-energy surface is represented as a function of the collective variables $d_1$ and $d_2$, defined as the distances between the carbon atom and the fluorine and chlorine atoms in \r{A}, respectively. Sampling points selected by BUQ are shown as white markers. The Minimum Free-Energy Path (MFEP), obtained via the string method, is shown in white. \textbf{b}, free energy along the MFEP and the structures of reactants, products, and transition state, indicating the reaction coordinates $d_1$ and $d_2$.}
\label{fig:sn2}
\end{figure}

Chemical-reaction free-energy landscapes are challenging; they are often rugged, high-dimensional, and reacting particles can move in an unbounded space~\cite{ensing2006,awasthi2019}. Moreover---even when using machine learning (ML) potential instead of density functional theory\cite{Behler2016}---the computation the atomic interactions at each timestep is order of magnitudes more costly than in classical MD. To further demonstrate the versatility of BUQ, we study the reaction free-energy landscape of an asymmetric S$_\text{N}2$ substitution between ethyl chloride and a fluoride anion. We use a machine-learned reactive potential \cite{Kuryla2025} and describe the reaction with two CVs, $d_1$ and $d_2$---i.e., the distances from the carbon to the fluorine and chlorine atoms, respectively (see Fig.~\ref{fig:sn2}). For each sampled point $(d_1,d_2)$, we measure the forces and feed them to our BUQ workflow. We then apply the string method~\cite{E2002,E2007} to the BUQ-predicted surface to extract the minimum free-energy path (MFEP) (for more details see the Supporting Information). 

In Fig. \ref{fig:sn2}, we show BUQ sampling results, including the whole free-energy landscape and the reaction barrier. BUQ was performed with a Matérn-$5/2$ kernel and a lengthscale of 0.2 \r{A} in both dimensions, and a weight exploiting the free-energy minima $\lambda = 0.1$. After 3 initial samples and 47 BUQ queries, the predicted free-energy barrier shows convergence (see Supporting Information). The MFEP projected onto $(d_1.d_2)$ agrees well with the minimum potential-energy path (MEP) reported in \cite{Kuryla2025}, with only small differences due to the widening of the free-energy minima at finite temperature. The reactants-to-products free-energy barrier of 0.044 eV is lower that the reported potential-energy barrier of 0.261 eV. This difference between the MFEP and the MEP barried could be attributed to entropic contributions. However, other free-energy calculations have reported barriers closer to 0.3 eV, ~\cite{pliego2011chemical,li2022atomic} indicating that perhaps the ML potential requires additional refinement. Still, when considering this interaction potential, BUQ quickly converges and maps out the free-energy surface using few samples.

\section{Discussion}




We have developed Bayesian Umbrella Quadrature (BUQ), a more efficient and universal free-energy calculation method combining the well-known Umbrella Integration (UI) framework with Bayesian quadrature (BQ). BUQ overcomes two of the largest challenges in standard UI: (1) interpolating the free energy between gradient samples and (2) selecting where to place these samples. In our approach, the free-energy gradient is modeled by a Gaussian Process (GP)---i.e., a noise-tolerant probabilistic model---while an acquisition function iteratively proposes the most informative gradient samples to reduce the variance of the integral, i.e., the free energy. In other words, instead of uniformly covering the CV-space with numerous windows, we iteratively choose the next bias location that maximally improves our estimate of the free energy. To our knowledge, this is the very first application of BUQ to enhanced sampling MD simulations. Our tests show that this adaptive BUQ strategy converges significantly faster than standard, i.e. uniform, UI. This is first shown in the $(\phi,\psi)$ conformational free-energy calculation of alanine dipeptide, where, to reach 1 kcal/mol accuracy, we see a 1.6 sample-efficiency speedup compared to a 10×10 standard umbrella grid (Fig.~\ref{fig:adipep}). This speedup can be explained by the two aforementioned characteristics of BUQ. On one hand, the GP provides a smooth model of the free-energy gradient that can interpolate between samples. Prior studies have shown that GP-based reconstruction can greatly reduce cost by exploiting correlations in the free-energy data~\cite{Ladygin2021,Mones2016,Stecher2014}. On the other hand, the acquisition function effectively samples key regions---for example, near barriers or changes in curvature. Our results confirm that a Bayesian sampling strategy can improve MD-based free-energy calculations, in agreement with the broader literature on probabilistic numerics~\cite{kanagawa2019,gessner2020}. 

To demonstrate that the efficiency of BUQ extends across systems and processes, we apply to method to two additional, fundamentally different, rare events. First, we tackle the phase-change from water to ice Ih, described by the number of ice-like water molecules, $N_\text{ice}$, as determined by an environment similarity order parameter. BUQ performance shines again. Compared to variational enhanced sampling, it delivers a 2.8 speedup considering the total number of MD timesteps. The nucleation free-energy barriers agree within 0.1 $NkT$, a small discrepancy attributable to the difference between static and dynamic biasing methods. Second, we study an asymmetric $S_N2$ chemical reaction between ethyl chloride and a fluoride anion. Even when using an ML potential, computing each new timestep can be extremely more expensive than when using classical forcefields. Here, BUQ invests only 50 samples to compute a reaction free-energy landscape, which validates well against convergence checks and related literature. Our results show how BUQ can become a general tool across different domains and phenomena, including structural transitions, phase equilibria, and reaction pathways.

One practical challenge when deploying BUQ is choosing its hyperparameters, such as the type of kernel or lengthscale for the GP \cite{Rasmussen2006}. First, we note that similar requirements arise in all enhanced sampling methods, such as the choice of the Gaussian width, height and deposition frequency in metadynamics\cite{laio2002,Bussi2020}. One advantage of BUQ is that the collected data points can always be recast using a different GP setup without rerunning biased MD simulations, so the free energy can be easily reinterpreted. Moreover, many choices of BUQ hyperparameters can lead to successful free-energy calculations, as we show in the Supporting Information, indicating that the method is robust. Here, our BUQ runs were tuned manually, we defer a more systematic hyperparameter study to future work. Another limitation of all enhanced sampling methods is diagnosing convergence; because BUQ provides a posterior distribution on the free energy, we have a measure of uncertainty, but turning that into a clear stopping criterion is nontrivial. In practice one must still inspect the evolving free-energy surface for convergence. Same as in standard UI, one can use error analysis to monitor the unimodality and consistency of the biasing force. Also like in standard UI, steering the system to a target window can occasionally fail if the bias is too weak to overcome a barrier, or if the steering occurs to quickly. Finally, BUQ, same as UI or US, assumes that the CVs are well-chosen; as always, if important motions lie outside the chosen variables, the free energy will not be captured accurately. 

Looking forward, there are many BUQ extensions to explore in future work. In this study, we focused on getting a reliable free-energy landscape overall. However, with small adaptations to the acquisition function, BUQ can also be used to focus on stable basins or transition states. BUQ's acquisition function could also be used to, for example, guide the refinement of an ML potential at regions when one wishes to improve a free-energy estimate. Moreover, multi-fidelity Bayesian strategies~\cite{Sabanza-Gil2025} offer an avenue to compute free energies at varying accuracies and computational costs. When looking at the GP, one can explore many more types of kernels, including quantum ones~\cite{Dai2023,rapp2024}, whose expressivity might be better suited for certain free-energy landscapes. BUQ's probabilistic framework can also be extended to multiple related integrals~\cite{Xi2018}, which could be useful when studying similar molecules or materials. For example, one could simultaneously compute free-energy surfaces for the same system at different temperature or pressure conditions, accelerating high-throughput screening. Moreover, a non-harmonic functional form of the bias could also be evaluated, such as Gaussians or polynomials, enabling exploration with a single simulation and possibly leading to even faster convergence of the free energy. Finally, beyond the three types of rare events showcased here, our open-source BUQ framework eases the application to further transitions, such as alchemical or binding free-energy calculations with appropriate CVs. In summary, by demonstrating faster convergence and broad applicability, our BUQ pipeline opens the door to a new generation of automated free-energy methods that combine enhanced sampling and Bayesian inference.

\begin{acknowledgement}

This work used the Dutch national e-infrastructure with the support of the SURF Cooperative using grant no. EINF-10854. Computations were performed using the University of Amsterdam - Science Faculty (UvA/FNWI) High Performance Computing Facility, a centrally managed computational resource available to UvA/FNWI researchers including faculty, staff, students, and collaborators. This publication is part of the project ``PLASTIC-JUNC: Predicting nanoPLASTIC risks to animal JUNCtional protein integrity'' with file number VI.Veni.232.120 of the research programme ``NWO Talent Programme 2023'' which is (partly) financed by the Dutch Research Council (NWO).
\end{acknowledgement}

\section*{Code and Data availability}
Code to reproduce all analysis and results is available at: \\ \url{https://github.com/computational-chemistry-uva/BUQ}.

\section*{Author contributions statement}

A.P.A.O. conceptualized the idea and the test cases. Both authors designed the project. E.K.K. produced all code to generate, pre- and post-process simulations, conducted the simulations, analyzed and validated the results together with A.P.A.O., produced visualizations, and wrote the manuscript. Both authors reviewed the manuscript. 

\section*{Additional information}
The authors declare no competing interests.
\section{Supporting Information}

\section{Gaussian Process Surrogate Models}

The choice of covariance kernel determines the prior smoothness assumptions about the mean biasing force, i.e., the free-energy gradient. In this study, we have mainly used three different kernels: the Radial Basis Function (RBF), Matérn-$1/2$ and Matérn-$5/2$~\cite{Rasmussen2006}. We employ stationary kernels with automatic relevance determination (ARD), allowing each collective variable (CV) dimension to have its own characteristic lengthscale. On top of these three kernels, it is possible to add a a white-noise term to represent measurement noise originating from finite sampling in the biased Molecular Dynamics (MD) simulations. In this section, we will shortly discuss the different kernels.

\subsection{Radial Basis Function (RBF) Kernel}
The RBF, also known as squared–exponential kernel assumes an infinitely differentiable underlying function. For CVs $\mathbf{s},\mathbf{s}^\prime \in \mathbb{R}^d$,
\begin{equation}
k_{\mathrm{RBF}}(\mathbf{s},\mathbf{s}') = \sigma^2 \exp\left(-\frac{1}{2}\sum_{j=1}^d \frac{(s_j-s'j)^2}{\ell_j^2}\right).
\end{equation}
Here, the hyperparameters are $\ell_j$ ,the lengthscale in dimension $j$, and $\sigma^2$, the process variance. The RBF kernel imposes strong smoothness, the integrand and all derivatives exist and are continuous. When added to a white-noise component
\begin{equation}
k_{\mathrm{white}}(\mathbf{s},\mathbf{s}') = \tau^2\delta_{\mathbf{s}\mathbf{s}'},
\end{equation}
the full kernel becomes $k = k_{\mathrm{RBF}} + k_{\mathrm{white}}$. When setting the noise-parameter $\tau = 0$, it reduces to just an RBF kernel.  

\subsection{Matérn Kernels}
The Matérn family provides controlled smoothness via the parameter $\nu$. The Matérn kernel reduces to RBF in the limit $\nu \to \infty$. Using ARD lengthscales with Euclidean metric, we define
\begin{equation}\label{eq:r}
r(\mathbf{s},\mathbf{s}') = \sqrt{\sum_{j=1}^d \frac{(s_j-s'j)^2}{\ell_j^2}}.
\end{equation}
The full Matérn kernel with smoothness parameter $\nu$ is given by
\begin{equation}
k_{\mathrm{M}\nu}(\mathbf{s},\mathbf{s}')
=
\sigma^2 \frac{2^{1-\nu}}{\Gamma(\nu)}
\left(\sqrt{2\nu}\, r\right)^{\nu}
K_{\nu}\!\left(\sqrt{2\nu}\, r\right),
\end{equation}
where r is defined in Eq.~\ref{eq:r}, $\sigma$ is the variance, $\Gamma(\cdot)$ denotes the Gamma function, and $K_\nu(\cdot)$ is the modified Bessel function of the second kind.
The Matérn-$1/2$ kernel describes non-differentiable functions and is equivalent to the exponential kernel:
\begin{equation}
k_{\mathrm{M12}}(\mathbf{s},\mathbf{s}') = \sigma^2 \exp(-r),
\end{equation}
with $r$ defined as in Eq.~\ref{eq:r} and $\sigma$ is again the variance. This kernel sets $\nu = 1/2$ and imposes minimal smoothness, which allows for sharp variation in the mean force profile. In contrast, the Matérn-$5/2$ kernel (i.e. with $\nu = 5/2$) corresponds to twice differentiable sample paths. It is defined as:

\begin{equation}
k_{\mathrm{M52}}(\mathbf{s},\mathbf{s}') = \sigma^2\left(1 + \sqrt{5}r + \frac{5}{3}r^2\right)\exp(-\sqrt{5}r),
\end{equation}
with $r$ defined as in Eq.~\ref{eq:r} and $\sigma$ is the variance. The Matérn-$5/2$ kernel is suitable when the mean force is smoother than once-differentiable fields but not as smooth as RBF. Equivalent to the RBF, we can also add the white kernel to represent the noise from the MD sampling, so the full model uses $k = k_{\mathrm{M}} + k_{\mathrm{white}}$. 

\subsection{Practical Implementation}
In all cases, the GP is constructed with ARD kernels over a maximum of two CV dimensions, where a white-noise term can be added for measurement uncertainty. Kernel selection controls the prior regularity of the free-energy gradient (i.e., mean biasing force). RBF favors globally smooth surfaces, while Matérn variants allow increasingly rough behavior. All kernels are defined using the GPY library~\cite{GPy2012} and then wrapped in the Emukit~\cite{Paleyes2019,Paleyes2023} library. For all Emukit kernels, we use the Lebesgue measure, which is a uniform distribution.

\section{Acquisition Functions}
Acquisition functions guide the sequential selection of new sampling locations in Bayesian Quadrature (BQ) ~\cite{gessner2020, gunter2014}. At each iteration, the next CV point is chosen by maximizing the acquisition that quantifies the expected utility of evaluating the mean force at that location. All strategies considered here depend directly on the GP posterior. We describe two classes: uncertainty sampling and integral–variance–reduction (IVR), as well as the free-energy minima exploiting term used in the main text.

\subsection{Uncertainty Sampling}
Uncertainty sampling selects points where the GP predictive variance of the integrand is maximal ~\cite{gunter2014}. For a GP surrogate of $-\mathbf{F}(\mathbf{s})$ with posterior variance $k_D(\mathbf{s},\mathbf{s})$, the acquisition is simply

\begin{equation}
a_{\rm US}(\mathbf{s}) = k(\mathbf{s},\mathbf{s})  \cdot (p(x))^q.
\end{equation}

\noindent Here, $p(\mathbf{s})$ is the density of the integration measure, and $q$ is the measure power parameter. The default choice in Emukit is $q=2$. When the integration measure is uniform (Lebesgue), it reduces to:
\begin{equation}
a_{\rm US}(\mathbf{s}) \propto  k(\mathbf{s},\mathbf{s}),
\end{equation}
reducing to the classical predictive–variance criterion widely used in uncertainty sampling. 

This prioritizes locations where the surrogate of the mean force is least certain, independent of their impact on the integral. Uncertainty sampling is computationally inexpensive and provides broad exploration when little is known about the landscape. In quadrature problems, it can oversample regions irrelevant for the integral, motivating criteria that consider both predictive variance and relevance to the integral estimate. However, since it is economical and easily implemented, we tested it as well. 

\subsection*{Integral Variance Reduction (IVR)}
The IVR strategy selects points expected to maximally reduce the posterior variance of the integral.~\cite{gessner2020} The acquisition is the expected drop in integral uncertainty when augmenting the dataset with a new observation:
\begin{equation}
a_{\rm IVR}(\mathbf{s}) = \mathbb{V}n - \mathbb{V}\bigl[A \mid D \cup \mathbf{s}\bigr].
\end{equation}
Here, $\mathbb{V}n$ is the current variance of $A$, and $\mathbb{V}\bigl[A\mid D\cup{\mathbf{s}}\bigr]$ is the variance when the next observation $\mathbf{s}$ is sampled.
This may also be written as the squared correlation $\rho^2(\mathbf{s})$ between the integrand value at $\mathbf{s}$ and the integral. IVR thus focuses sampling in regions that most influence the quantity of interest rather than simply uncertain locations.

\subsection{Exploitation of free-energy minima}
To accelerate the finding of minimum free–energy pathways, we combine IVR or uncertainty sampling with a term that directs sampling toward predicted low–free–energy regions. If $a_{\rm FES}(\mathbf{s})$ denotes the normalized GP estimate of $A(\mathbf{s})$ and $\lambda\in[0,1]$ controls trade–off,
\begin{equation}
a_{\rm total}(\mathbf{s}) = -\lambda a_{\rm FES}(\mathbf{s}) + (1-\lambda)a_{\rm IVR | US}(\mathbf{s}).
\end{equation}
Here, both $a_{\mathrm{IVR |US}}$ and the current FES estimate $a_{\rm FES}$ are normalized. The first term favors exploitation of likely low–free–energy states, while the second enforces global exploration via variance reduction of the integral (IVR) or variance reduction of the force profile (Uncertainty Sampling). For $\lambda=0$, the strategy reduces to pure IVR or pure uncertainty sampling, whereas $\lambda\rightarrow 1$ favors pure exploitation. This combined approach suppresses unnecessary sampling in high–free–energy regions and accelerates refinement in physically relevant regions.

\section{Conformational change in alanine dipeptide}

\subsection{Molecular Dynamics protocol}\label{sec:simproc}
To establish a benchmark, we carried out simulations of alanine dipeptide in vacuum. An initial unbiased molecular dynamics trajectory of 5~ns was done with GROMACS 2022.5~\cite{Berendsen1995}, to equilibrate the system. The system was simulated at 300~K using the AMBER99SB-ILDN~\cite{LindorffLarsen2010} force field and the Verlet cutoff scheme with PME electrostatics (real-space tolerance $1\times10^{-5}$; Lennard-Jones tolerance 0.001). The Coulomb cutoff and the van der Waals cutoff were both set to 1 nm.  The cubic box had an edge length of 2.72~nm.
Bond constraints on all hydrogen-involving interactions were imposed using LINCS \cite{Hess1997}.
The integration timestep was 2~fs, and temperature was controlled with the  velocity rescale thermostat with a time constant of  0.1~ps \cite{Bussi2007}. For the rest of the discussion of Alanine Dipeptide, the same settings are used, unless specified otherwise. The result of this 5 ns unbiased MD will be used as a starting point for the metadynamics and for the umbrella sampling runs. 

\subsection{Metadynamics protocol}
To get a benchmark, we performed a 100~ns metadynamics simulation in Gromacs 2023 using plumed 2.9.1~\cite{Tribello2014,consortium2019}. As CVs, we used the backbone dihedral angles $\phi$ and $\psi$, with Gaussian potentials deposited every 2~ps. The Gaussian height was 0.1~kcal/mol with a width of 0.25~rad in each CV. All PLUMED input files are provided in the repository 

\begin{figure}
        \centering
        \includegraphics[width=\linewidth]{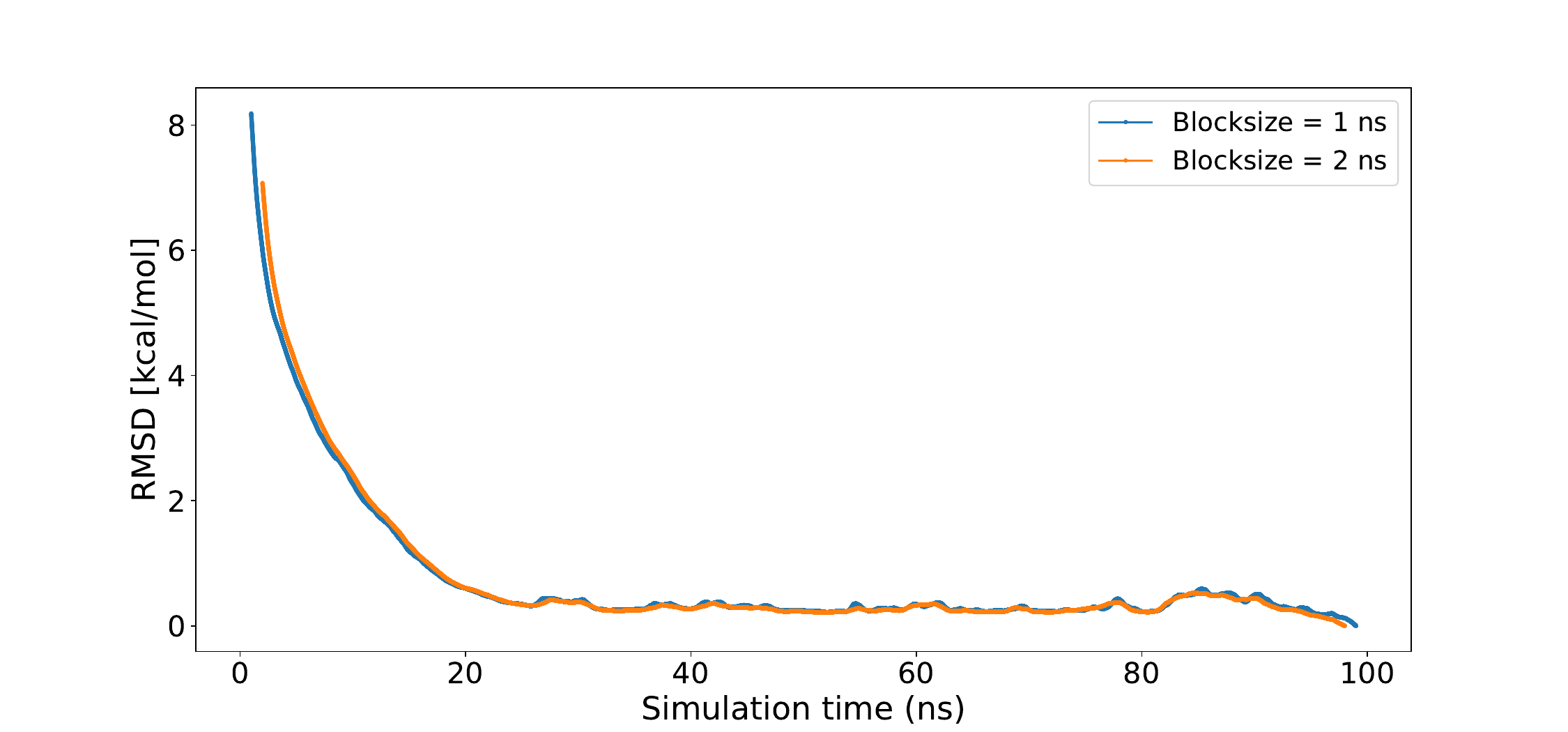}
        \caption{Block-averaging convergence analysis of the benchmark metadynamics free energy surface (FES). FES estimations were averaged over blocks of 500 (1 ns) and 1000 (2 ns) Gaussian depositions, aligned to zero at their minimum, and compared to the final block via RMSD. }
        \label{fig:conf}
    \end{figure}

To assess the convergence of the benchmark metadynamics simulation, we analyzed the generated free energy surfaces (FES), after the deposition of each Gaussian. To assess convergence, a
block-averaging scheme was employed, using blocks of 500 and 1000 FES files (1~ns and 2~ns, respectively). For each block, the free energies were averaged and aligned to zero at their minimum. Convergence was quantified via the root-mean-square deviation (RMSD) relative to the final block, see Fig.~\ref{fig:conf}. For the last 50 ns, the largest deviation of the FES from the final 2 ns (the final FES-block average), is 0.52 kcal/mol. This block-average of the final 2 ns is taken as the benchmark of the Bayesian quadrature results.

\subsection{Umbrella Sampling protocol}
For the US and BUQ samples, we run short restrained MD at different target values of $(\phi^\star,\psi^\star)$. These biased simulations are done with GROMACS 2023~\cite{Berendsen1995} patched with PLUMED 2.9.1~\cite{Tribello2014}. The simulation time was 0.2 ns per sample. All MD simulation settings are as described in~\hyperref{sec:simproc}, and all samples are started from the equilibrated state. PLUMED input files declared
\texttt{UNITS LENGTH=A TIME=ps ENERGY=kcal/mol} and defined the backbone torsions
$\phi$ and $\psi$ together with their trigonometric components $\{\sin\phi,\cos\phi,\sin\psi,\cos\psi\}$, which ease the biasing or periodic CVs.
 \texttt{MOVINGRESTRAINT} harmonic biases were applied on the four trigonometric CVs with the same force constant
$\kappa = 200$ kcal/mol. The biasing schedule is as follows:
\begin{enumerate}
\item an unbiased start (500 timesteps)
\item gradual increase of the force constant up to the target $\kappa$ by step 1500
\item smooth translation of the bias centers from the initial values $(\phi_0,\psi_0)=(-1.580543,\,1.311767)$ to the target window $(\phi^\star,\psi^\star)$ at a speed proportional to the change in angles, $|\Delta\theta|$, completed by step $1500+1000\,|\Delta\theta|$
\end{enumerate}

Each query consisted of $N_\text{steps}=100{,}000$ MD timesteps (200~ps total), and we used the last 100~ps for measurement of the biasing force. PLUMED printed $\sin\phi$, $\cos\phi$, $\sin\psi$, $\cos\psi$ every 100 steps to a file. Convergence of the force was verified via block analysis.

From the time-averaged trigonometric CVs $(\overline{\sin\theta},\overline{\cos\theta})$ and the restraint targets $(\sin\theta^\star,\cos\theta^\star)$ for $\theta\in\{\phi,\psi\}$, we formed the two harmonic forces
$f_{\sin\theta}=k_\theta(\overline{\sin\theta}-\sin\theta^\star)$ and
$f_{\cos\theta}=k_\theta(\overline{\cos\theta}-\cos\theta^\star)$ and projected them onto the angular direction using the phase difference via $\operatorname{atan2}$:
\[
F_\theta = -\,\operatorname{sgn}\!\Big(\operatorname{atan2}(\overline{\sin\theta},\overline{\cos\theta})
- \operatorname{atan2}(\sin\theta^\star,\cos\theta^\star)\Big)\,
\sqrt{f_{\sin\theta}^2+f_{\cos\theta}^2},
\]
yielding the desired noisy estimate of $(-\partial_\theta A)$ at $(\phi^\star,\psi^\star)$.

\subsection{Obtaining and benchmarking a free-energy landscape from BUQ}

In these simulations, angles were treated as periodic with domain $\phi,\psi \in [-\pi,\pi)$, and all energies are in kcal/mol, times in ps, angles in radians. 
To obtain a free-energy surface from BUQ, we modeled the 2D gradient field with a Gaussian process using \textbf{GPy}~\cite{GPy2012} and \textbf{Emukit}~\cite{Paleyes2019,Paleyes2023}. Unless otherwise noted, kernels were stationary and used ARD; a small diagonal white kernel component could be added to model measurement noise. The GP was wrapped in Emukit’s quadrature interface with a Lebesgue prior over $[-\pi,\pi]^2$ and trained on the acquired pairs $\{(\phi_i,\psi_i),(-F_{\phi,i},-F_{\psi,i})\}$. We then predicted the gradient on a $100\times 100$ grid and integrated it to a free energy using a composite trapezoid rule on the grid, followed by an L-BFGS-B refinement that minimizes the squared mismatch between the numerical gradient of the surface and the GP-predicted gradient~\cite{boneta2021umbrella}. The resulting free-energy surface was shifted to have minimum zero. Accuracy was quantified by the Root Mean Squared Deviation (RMSD) with respect to the ground-truth metadynamics reference, projected onto the same grid points:

\begin{equation}
    \mathrm{RMSD} \;=\; \sqrt{\frac{1}{10000}\sum_{i=1}^{100}\sum_{j=1}^{100}\Big( A_{\rm prediction}(s_{ij}) - A_{\rm metadynamics}(s_{ij})\Big)^2 }.
\end{equation}
The same procedure is done to obtain and benchmark free-energy surfaces from the standard UI simulations with uniform grids. In these cases, the pairs $\{(\phi_i,\psi_i),(F_{\phi,i},F_{\psi,i})\}$ are determined by the uniform grid. 

Here we provide details about both cases:
\begin{enumerate}
\item   BUQ: We initialized with two samples at the minima of the surface, $(\phi,\psi) =(-1.508,\,0.880)$ and $(1.194,\,-0.880)$ and updated the GP after each new measurement. We evaluated two acquisitions:
  Integral Variance Reduction (IVR) and Uncertainty Sampling (US). We have used this with both RBF and Matérn-$5/2$ kernels, varying the lengthscale and the minima exploitation parameter $\lambda$ (see below). We repeated this sampling for a fixed budget of 98 queries, i.e., 100 samples in total. 
  \item Standard UI on uniform grid: We queried an $n\times n$ lattice with $n=1,\dots,10$ (i.e., from 1 to 100 grid points). We used an RBF kernel with lengthscale $\ell=0.5$ and no additional white noise. For fairness, each query used the same MD budget described above. RMSD was reported as a function of $n^2$ queries. To consider periodicity, points sampled exactly at a dihedral angle of $-\pi$ were duplicated on the opposite boundary.

\end{enumerate}

\subsection{Bayesian Umbrella Quadrature protocol}
To assess hypermarameters, we performed several BUQ runs, where we varied the lengthscale $\ell\in\{0.5,0.6\}$ rad, for the RBF kernel , and we set  $\ell = 0.75$ rad for the Mat\'ern-$5/2$ kernel. We also implement white-noise variance $\in\{0,0.1\}$ (only for the Mat\'ern-$5/2$) and acquisition $\in\{\text{IVR},\text{US}\}$, (for both kernel types) and $\lambda\in\{0,0.1,0.2\}$ (for both kernel types), all with 100 queries. These settings where decided after an initial exploration. The best-performing configuration (lowest RMSD after 100 queries) was: Mat\'ern-$5/2$ kernel (ARD) with lengthscales $\ell=0.75$, zero added white noise, IVR acquisition with $\lambda=0.1$, and 100 sequential queries. This setting was used for the BUQ results in the main text. 

\begin{figure}
        \centering
        \includegraphics[width=1.\linewidth]{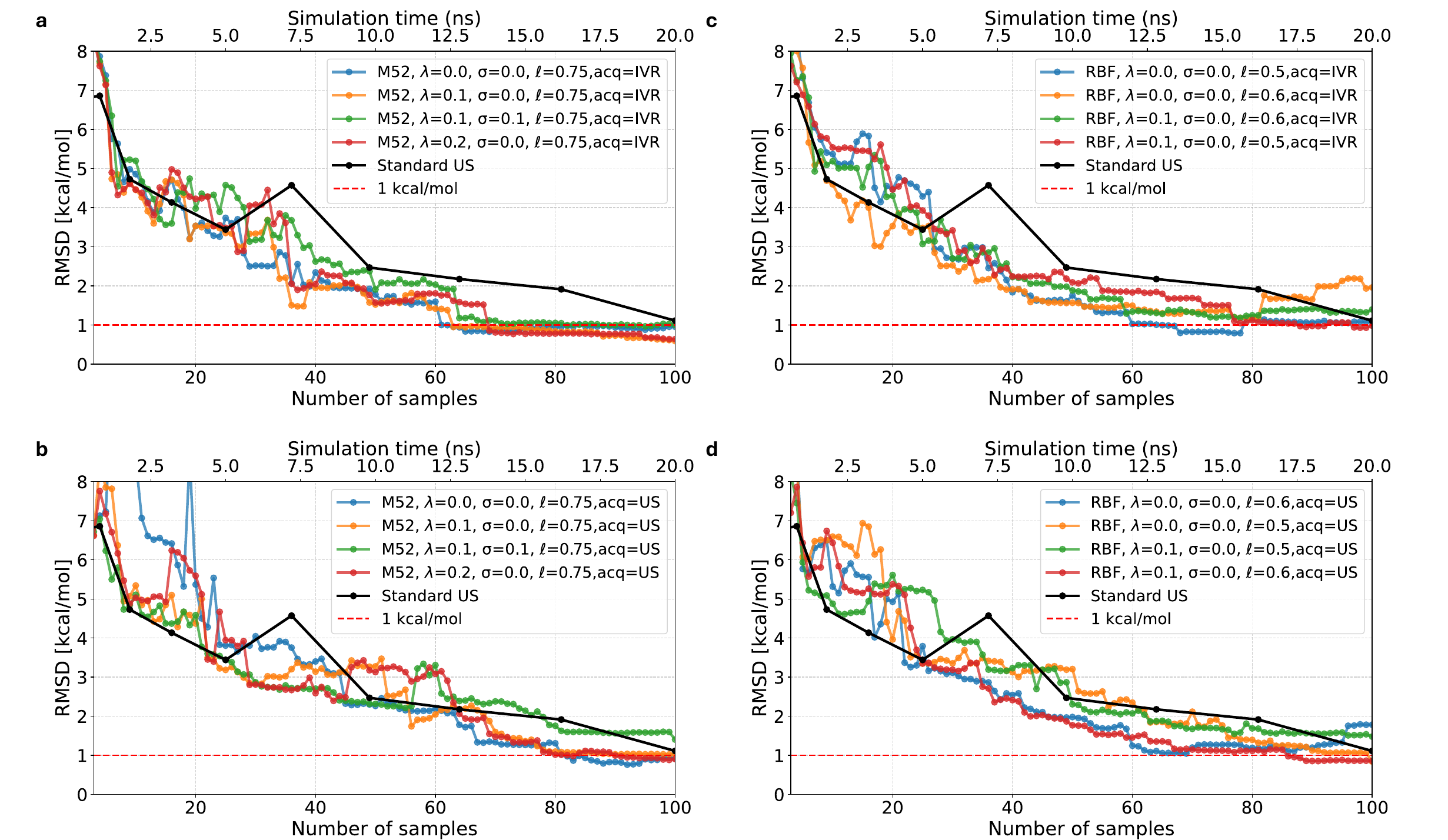}
        \caption{Free-energy landscape root-mean-square deviation (RMSD) with respect to the metadynamics ground truth as a function of the number of Bayesian umbrella quadrature (BUQ) queries with different hyperparameters. (a) Mat\'ern-$5/2$ kernel with integral variance reduction (IVR) acquisition. (b) Mat\'ern-$5/2$ kernel with uncertainty sampling acquisition. (c) RBF kernel with IVR acquisition. (d) RBF kernel with uncertainty acquisition. For the RBF kernel, the lengthscale was varied as $\ell\in\{0.5,0.6\} $ rad, while for the Mat\'ern-$5/2$ kernel $\ell=0.75$ rad was used. The white-noise variance was $\sigma\in\{0,0.1\}$ for Mat\'ern-$5/2$ only. The exploitation parameter was $\lambda\in\{0,0.1,0.2\}$ for both kernel types. Each configuration was evaluated with 2 initial samples followed by 98 acquired samples. Black curves denote standard umbrella sampling (US) on a uniform grid, and the red dashed line indicates the 1 kcal/mol target accuracy. }
        \label{fig:kernelhyper}
\end{figure}

All Python scripts, PLUMED templates, and COLVAR files used to generate and post-process the BUQ training data (including the hyperparameter sweep) are provided in the repository. All restrained MD simulations were performed with GROMACS 2023 patched with PLUMED 2.9.1 using \texttt{UNITS LENGTH=A TIME=ps ENERGY=kcal/mol}.

\section{Phase transition from water to ice}

\subsection{Molecular Dynamics protocol.}
All simulations were carried out with LAMMPS (29 Aug 2024)~\cite{Thompson2022} (units = \texttt{real}) patched with PLUMED 2.9.1~\cite{Tribello2014}, using the TIP4P water model~\cite{Abascal2005}. The simulation box contained 864 water molecules, initialized from an equilibrated ice configuration. MD were performed with a 2 fs timestep using velocity rescaling~\cite{Bussi2007} at 300 K with a 100 fs time constant, and an isotropic NPH barostat~\cite{shinoda2004rapid} at 1 bar with a 1 ps  time constant, and SHAKE~\cite{Krutler2001} constraints applied to O–H bonds and H–O–H angles. This setup and LAMMPS input files are taken from Ref. \cite{Piaggi2020}.

\subsection{Umbrella Sampling protocol}

The CV, the environment similarity order parameter~\cite{piaggi2019}, of this simulation is defined as follows:
Two environment similarities were defined, comparing instantaneous oxygen environments to reference ice structures (hexagonal and cubic). These descriptors yielded scalar, atom-specific order parameters $q_\mathrm{Ih}$ and $q_\mathrm{Ic}$.
To avoid competition between the two ice phases, we restrain the normalized growth of the system-averaged $\bar{q}_\mathrm{Ic}$ to be less than that of the target $\bar{q}_\mathrm{Ih}$. For more details, see \cite{Piaggi2020}.
In practice, we bias the number of Ih ice-like oxygens $N_\text{ice}$, with an $q_\mathrm{Ih}$ value higher than 0.5. 

All queries started from an ice system, i.e., $N_{\text{ice}}=$ 288. The biasing schedule is similar to the alanine dipeptide case. First, the $\kappa$ is gradually built up from 0 to 100 kj per particle in 1000 timesteps. Then, it is moved to the desired window target in 49000 timesteps. We simulate for 15 ns in total, and we use the last 14 ns for statistics. CV values were stored every 500 steps. All PLUMED files are available in our repository. 

\subsection{Bayesian umbrella quadrature protocol}

To initialize, we take four samples near the stable states, with $N_{\text{ice}}$ values of 1.6, 3.0, 284.16 and 285.0 to correctly capture the liquid and crystal minima. These values are chosen based on a histogram analysis of the value of $N_{\text{ice}}$ of unbiased MD simulations near the stable states. After these four initial points, we use BUQ with IVR acquisition to obtain 15 more queries, resulting in 19 simulations in total. The baseline for accuracy is the reference free-energy profile reported in Ref.~\cite{Piaggi2020}.

\begin{figure}
    \centering
    \includegraphics[width=\linewidth]{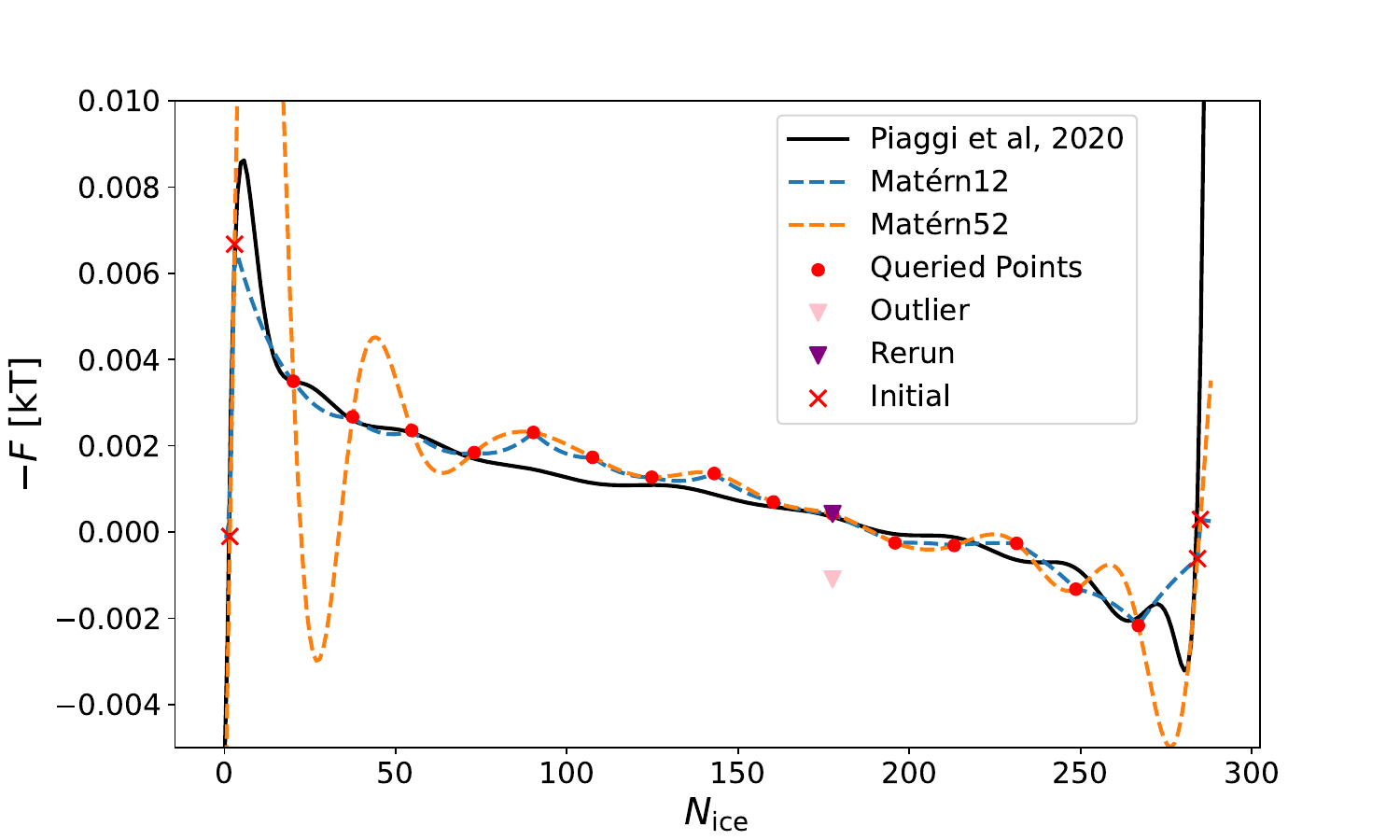}
\caption{Free-energy derivative $-F$ as a function of $N_{\text{ice}}$ for the ice–liquid phase transition. The black solid line shows the reference profile from Ref.~\cite{Piaggi2020}, while the dashed curves correspond to Gaussian process reconstructions with Matérn-$1/2$ and Matérn-$5/2$ kernels. Red indicates the 19 queried simulation points obtained through BUQ with IVR acquisition, starting from four initial samples (indicated by a red cross) near the stable liquid and crystal states. The pink inverted triangle marks a rejected outlier sample, attributed to the formation of an ice cylinder across periodic boundaries, and the purple inverted triangle denotes the rerun at the same $N_{\text{ice}}$. The Matérn-$1/2$ kernel successfully captures the ragged, non-smooth structure of the free-energy gradient associated with nucleation and melting.}
    \label{fig:derivative}
\end{figure}

Unlike the alanine-dipeptide system, the free-energy gradient of this phase transition is a more ragged and non-smooth function, reflecting nucleation and melting. To model this, we restricted the kernel family to Matérn-$1/2$ , which is able to capture sharp changes in the field, see also Fig.~\ref{fig:derivative}. The GP was wrapped in Emukit’s quadrature interface with a Lebesgue prior over the ES domain, trained on pairs $\{N_\text{ice},_i,-F_i\}$, and integrated via trapezoidal rule to yield a free-energy profile, shown in the main text.

We tested a small hyperparameter grid. Force constant $\kappa_\text{ice}$ was fixed at 100 kj per particle throughout production runs. An adaptive-$\kappa$ variant was evaluated, using a higher $\kappa$  in regions where a steeper derivative was predicted. However, for this system we did not see an effect, perhaps because the base $\kappa$ value was already strong enough to drive nucleation.  The kernel type was Matérn-1/2, known to handle non-smoothness more effectively than RBF, with lengthscales $\ell \in \{10,20\}$ $N_\text{ice}$.  The Gaussian white-noise variance was set to zero because additional noise obscured the steep gradients near the stable states, leading the GP to misinterpret them as stochastic fluctuations. Query budgets were $N_\mathrm{queries}=15$, with and without the exploitation acquisition term $\lambda \in \{0,0.1\}$.

After testing, a lengthscale of 20 $N_\text{ice}$ and $\lambda \in \{0.1\}$ delivered the best performance. However, we note that this run required one rerun due to a failed sample, which is indicated in Fig.~\ref{fig:derivative}. Since the sample appeared to be an outlier, even without access to the benchmark, we decided to rerun that point. This outlier is likely caused by the forming of an ice cylinder across the periodic boundaries of the box.  Moreover, this procedure shows how samples can easily be corrected within the BUQ framework. Only a simple refitting of the GP is required. 

\section{S$_\text{N}2$ chemical reaction between ethyl chloride and a fluoride anion}

\subsection{Molecular dynamics protocol}
We use a machine-learned reactive potential reported in Ref. \cite{Kuryla2025}. The MD simulations were carried out with ASE~\cite{HjorthLarsen2017}, and used a 0.5~fs timestep with a velocity rescaling thermostat at 300~K ~\cite{Bussi2007}. 

\subsection{Umbrella sampling protocol}
All simulations were initialized  with the ethyl initially bound to the chloride ion. We describe the reaction with two CVs, $d_1$ and $d_2$, i.e., the distances from the ethyl carbon to the fluorine and chlorine atoms, respectively.  Restraints were applied through PLUMED to enforce the desired geometries: the distances $d_1$ and $d_2$ were restrained with moving harmonic restraints that gradually steered the system from the starting geometry $(2.64,1.84)$~\AA{} to the target window $(d_1,d_2)$. the  Pulling along a reaction coordinate of the form $r_c = d_2 - d_1$ was avoided due to the difference in strength of the C–F bond and C-Cl bond. We used a $\kappa_{d_1} = 1000 $kcal/(mol$\times$ \AA{}) at 2.64 and a $\kappa_{d_2} = 100 $kcal/(mol $\times$ \AA{}) at 1.84 , which where both build up gradually in 1000 timesteps. Then, we used 4000 timesteps to move the restraint to the desired $d_1$ and $d_2$. Angular restraints were applied to prevent rotations of the reactive moiety, which could take the system into untrained configurations for the ML potential. CV values were stored every 100 steps. Each simulation ran for 80~ps, sufficient to equilibrate around the target geometry and compute average biasing forces on $d_1$ and $d_2$.  
§
\subsection{Bayesian umbrella quadrature protocol}
We considered the ranges $d_1 \in [1.2, 2.8]$~\AA{} and $d_2 \in [1.8, 3.5]$~\AA{}. We used the Matern52 kernel with a lengthscale of 0.2 \r{A} in both dimensions. Again, we set $\lambda=0.1$ for exploiting the minima, and we set the white noise to 0. We used three initial samples: two at the minima and one along the path connecting the minima, with values ($d_1, d_2)$ = (2.64, 1.84), (1.6, 2.6), (1.413, 3.468). Then, we used the IVR acquisition to obtain 37 more samples.

\begin{figure}
    \centering
    \includegraphics[width=\linewidth]{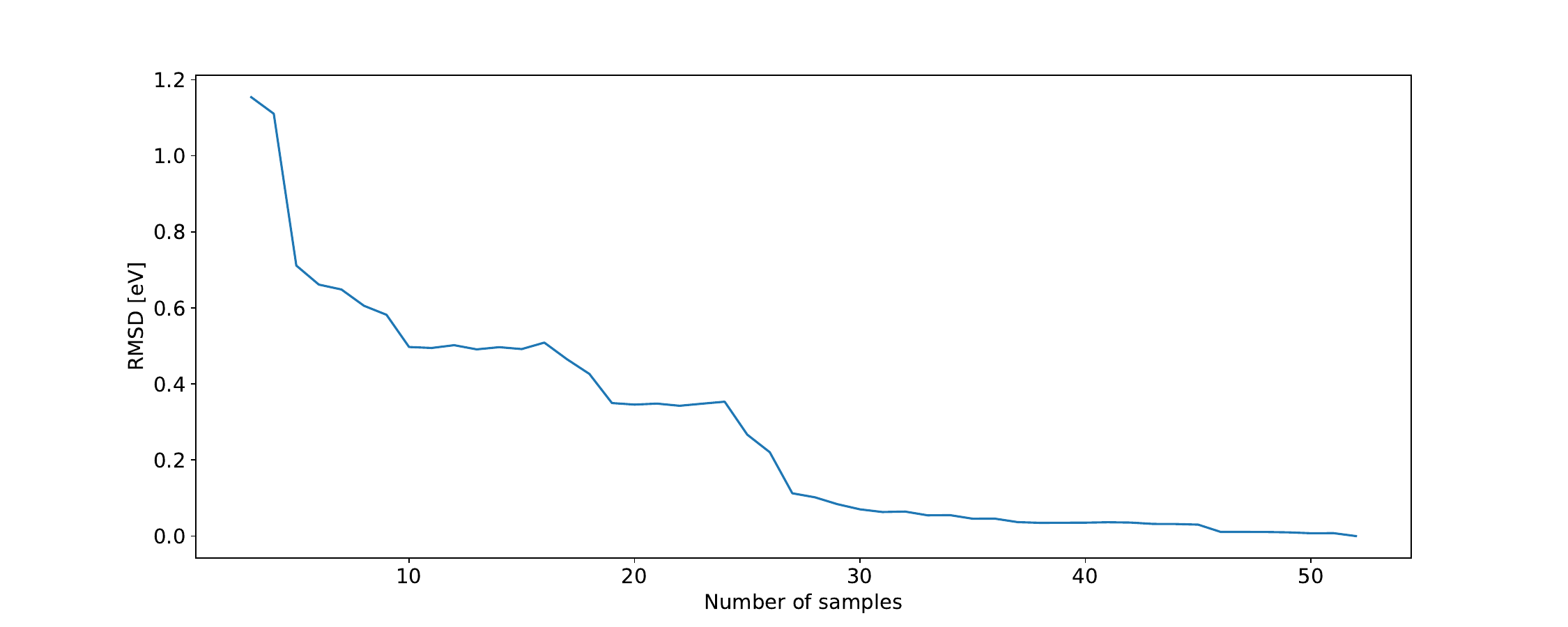}
\caption{Convergence of the free-energy surface with respect to the number of BUQ samples, measured by the RMSD between the current free-energy landscapes after $n$ samples and the final one after 53 samples in total., }
    \label{fig:convergencechemreaction}
\end{figure}

Finally, the BUQ-predicted surface was interpolated onto a 60$\times$60 grid, and a minimum free-energy path was extracted using the string method \cite{E2007}. The string was initialized as a straight-line path between the reactant (C–Cl bound) and product (C–F bound) states, discretized into 16 points. The path was relaxed using a fourth-order spline, with the endpoints fixed, over a maximum of 150 steps, and positions were recorded every 10 steps. The converged string reproduces the transition-state geometry reported in \cite{Kuryla2025}. The predicted free-energy barrier  matches the potential-energy barrier from the MACE potential. The valleys in the free-energy landscape are slightly wider than in the potential-energy surface, reflecting entropic contributions captured by the BUQ workflow. To further assess the convergence of BUQ, we performed an additional test by comparing the RMSD between the free-energy surfaces and the final free-energy landscape, see Fig.~\ref{fig:convergencechemreaction}. We start from the first 3 initial samples, and perform this analysis up to 50 additional samples (53 in total). In the end, we use the free-energy landscape after 50 samples in total, where the RMSD stabilizes below ~0.01 eV, indicating convergence.

\bibliography{final}

\end{document}